\documentclass[aps,showpacs,preprintnumbers,amsmath,epsf,amssymb,epsfig,nofootinbib]{revtex4}
\input epsf.sty
\usepackage{bm}
\begin{document}

\title{Inflation from the bang of a white hole
induced from a 6D vacuum state}
\author{$^{1}$Jos\'e Edgar Madriz Aguilar\footnote{
E-mail address: jemadriz@fisica.ufpb.br} and
$^2$Mauricio Bellini\footnote{E-mail address: mbellini@mdp.edu.ar}}

\address{$^1$ Departamento de F\'{\i}sica, Universidade
Federal da Para\'{\i}ba. C. Postal 5008, Jo\~{a}o Pessoa, PB 58059-970 Brazil.\\
$^2$ Consejo Nacional de Investigaciones
Cient\'{\i}ficas y T\'ecnicas (CONICET) and Departamento de
F\'{\i}sica, Facultad de Ciencias Exactas y Naturales, Universidad
Nacional de Mar del Plata, Funes 3350, (7600) Mar del Plata,
Argentina.}

\begin{abstract}
Using ideas of STM theory, but starting from a 6D vacuum state, we
propose an inflationary model where the universe emerges from the
blast of a white hole. Under this approach, the expansion is affected
by a geometrical deformation induced by the gravitational attraction of the
hole, which should be responsible for the $k_R$-non
invariant spectrum of galaxies (and likewise of the matter density)
today observed.
\end{abstract}


\maketitle
\section{Introduction}

In 1916 Karl Schwarzschild obtained a spherically symmetric
solution of Einstein field equations that we know as black hole %%@
solution. An interesting case is the
time reversal solution in which gravitational collapse to a black hole occurs.  %%@
This is known as the Schwarzschild white hole solution. In this framework a white hole is an object exploding from highly %%@
dense or singular state when it was originally well inside  its black hole.
Currently, this idea has been subject of great debate with
respect to its physical meaning, existence and importance. For some %%@
authors the physical relevance of such objects is considered rather
doubtful \cite{Novik}. However for some other researchers a %%@
current view of white holes bears to a revision of the standard cosmological
model (SCM) as it is done for example in %%@
\cite{temple}. In this new view is examined
the possibility that the big bang is a blast
that happened inside of a black hole,
followed by an expansion.
In other words, %%@
the possibility that our universe has been originated due to the explosion
of a white hole has already been considered in physics. %%@
As in general theory of relativity, some other %%@
objects that have similar properties to white holes have appeared in physics.
For instance, objects that have the unusual property %%@
of reflecting all test particles due to a repulsive (anti-gravitational) effect on them have been considered
in some treatments of %%@
Supersymmetry. These objects are called repulsons \cite{Repulsons}. Thus, in order to avoid misunderstanding, many researchers %%@
find convenient referring to white holes as compact objects which have the remarkable characteristic of having a repulsive %%@
naked singularity which reflects all test particles due to a repulsive (anti-gravitational) effect, being also this a better way %%@
to include all the possible configurations that could have these properties.
In this letter we will refer to a white hole %%@
in the former sense.\\

The idea that our universe is a 4D space-time embedded in a higher dimensional manifold with large extra %%@
dimensions has been a topic of increased interest in several branches of physics,
and in particular, in cosmology. This idea has  %%@
generated a new kind of cosmological models that includes quintessential expansion. In particular theories on %%@
which is considered only one extra dimension have become quite popular in the scientific community. Among these theories are %%@
counted the braneworld scenarios, the space-time-matter (STM) \cite{stm} and all noncompact Kaluza-Klein theories.
In this letter we have particular interest on the ideas of the STM %%@
theory. A success of the STM theory is that all matter fields in 4D can be geometrically induced from a 5D apparent
vacuum  leading an %%@
effective 4D energy-momentum tensor \cite{2} and for this reason the theory is also called Induced-Matter theory. The assertion %%@
that all matter fields can be geometrically induced is hardly supported by the Campbell-Magaard theorem and their
extensions %%@
\cite{CMT1}. In the case of braneworld cosmologies, our universe is modeled by a brane embedded in a higher dimensional manifold %%@
called the bulk. Matter fields are confined to the brane and gravity can propagate along the bulk. In general  braneworld models %%@
and STM  have different physical motivation and interpretation.
However, %%@
their equivalence has been recently shown by Ponce de Leon \cite{3}.
In both theories 4D physics is obtained by evaluating the 5D metric at some fixed hypersurface,
i.e. by establishing a foliation choosing a %%@
constant value of the fifth coordinate.
The possibility of having a dynamical foliation %%@
in a consistent manner was explored by the present authors in \cite{nintro1} and more
formally established by J. Ponce de Leon in %%@
\cite{nintro2}. In these papers by dynamical foliation the authors mean to take the fifth coordinate as a function of the cosmic %%@
time under the requirement of continuity of the metric, selecting in this way a dynamical 4D hypersurface. What makes relevant %%@
this new approach is that new physics can be derived from a new contribution of the fifth dimension in the induced matter on a %%@
time-varying 4D hypersurface.
This modification is absent when we choose a constant foliation \cite{nintro2}.\\

However,
there are some other questions that have not even been analyzed in the mentioned approach.
Is it possible to establish a dynamical foliation by choosing the fifth coordinate as a function of the rest of the spatial %%@
coordinates?
On the other hand, as it is well known the fifth dimension is considered in the STM theory as a geometric source of
the 4D physical %%@
matter fields. So that, as well as a dynamical foliation via a temporal dependence of the fifth coordinate allows
to  describe geometrically new sources of matter in 4D, is it possible to derive new physics by adding a spatial dependence in the %%@
fifth coordinate when implementing the foliation? Is there another mechanism to describe sources of matter by using foliations of
the higher dimensional space-time? In this letter we present a manner to address the problem outlined in the last question. We %%@
propose a new formalism where, instead of implementing  a dynamical foliation by taking a spatial dependence of the fifth %%@
coordinate including its time dependence, we consider another extra
dimension, the sixth dimension, in such a way that we
can implement two dynamical foliations in a sequential manner.
The first one by choosing the fifth coordinate depending of the cosmic time,
and the second one by choosing the sixth
coordinate dependent of the 3D spatial coordinates. All of these choices preserve
the continuity of
the metric. In addition, the 6D metric must be Ricci-flat. This
requirement is a natural extension of the vacuum condition
used in the STM theory, in which 5D Ricci-flat metrics are used. In simple
words, we are using the Campbell-Magaard
theorem and their extensions for embedding a 5D Ricci-flat
space-time in a 6D Ricci-flat space-time.
The conditions of 6D %%@
Ricci-flatness and the continuity of the metric give us
the foliation of the sixth coordinate. In other words, %%@
these conditions specify the sixth dimension as a function
of the 3D spatial coordinates, in order to establish the foliation.
This function, for a particular 6D metric, can be
seen in 4D as a gravitational potential associated to a localized
compact object %%@
that has the characteristics of a white hole.
From a more general
point of view, given a 6D Ricci-flat metric,
this is a mechanism %%@
for inducing localized matter onto a time-varying 4D
hypersurface by establishing a spatial foliation of a
sixth coordinate. This is very relevant because
we can describe matter at both, %%@
cosmological and astrophysical scales, in a cosmological model.
Of course, this is not the first 6D gravitational model developed %%@
from a 6D vacuum. The STM theory was extended  to more than five dimensions by Fukui \cite{PR1}. In his model, Fukui obtained a %%@
simple vacuum cosmological solution in the context of a Space-Time-Matter-Charge theory of the universe \cite{PR2}. This theory %%@
was proposed by Wesson in order to obtain a unified field theory of gravity and electromagnetism following the same line of his 5D %%@
STM theory \cite{PR3}.
On the other hand, a family of cosmological solutions to the Einstein
field equations obtained from a ($D+4$)-dimensional ($D \geq 2$) vacuum,
was studied in\cite{coley}.\\

Inflationary theory of the universe provides a physical mechanism to
generate primordial energy density fluctuations on cosmological scales \cite{infl}.
However, it fails when we try to predict the spectrum of
these fluctuations on more  smaller (astrophysical) scales.
Looking ahead for a novel cosmological approach that addresses this problem, we consider feasible to use the geometrical
formalism described previously to study an asymptotic
spatially flat FRW universe (on large scales) emerging from the
explosion of a white hole obtained from a 6D vacuum state.
We shall start our treatment considering a 6D vacuum
state rather than the usual 5D one \cite{n2}, to develop an effective 4D inflationary model of
the universe where the expansion is affected by a local geometrical deformation induced by the mass $M$
of a white hole induced in turn through the sixth dimension. Within this
approach, the 4D early universe
is viewed as a white hole that expands (the expansion is driven by
the scalar field $\varphi$) on the effective (4D) spatially curved
background metric, where the naked singularity is
identified with the big bang singularity.
Such a scenario should be feasible
on very extreme conditions, where the mass of the white hole is of
the order of the Planckian mass $M_p = 1.2 \times 10^{19}$ Gev, on
which Grand Unification mass scale could take place. A similar
approach (but without STM theory of gravity) was considered many
years ago in \cite{n1}. Since we are aimed to describe
an effective 4D cosmological scenario, the continuity condition %%@
applied to our 5D metric leaves a foliation on the fifth coordinate
in such a way that its time dependence is related to the Hubble %%@
parameter, while the foliation on the sixth coordinate induces a 4D
white hole gravitational potential. For simplicity, in this
letter will %%@
be considered a de Sitter expansion, where the Hubble parameter $H$
is a constant. However, our formalism could be extended for whatever $H=H(t)$.
Thus, in our model the fifth dimension is %%@
responsible for the 4D de
Sitter expansion, which is physically driven by the inflaton field
$\varphi$. From the
physical point of view, the sixth dimension
is responsible for the spatial curvature induced by the mass
of the white hole (located at $R=0$). In more general terms,
within our approach the fifth dimension is physically related to the
vacuum energy density which is the source of the effective 4D global inflationary expansion whereas the sixth one induces local %%@
gravitational sources.

\section{Effective 4D dynamics from a 6D vacuum state}

In order to describe a 6D vacuum, we consider the 6D Riemann flat metric
\begin{equation}\label{ini}
dS^2 = \psi^2 dN^2 - \psi^2 e^{2N} \left[dr^2+ r^2 d\Omega^2\right]
-d\psi^2 - d\sigma^2
\end{equation}
which defines a 6D vacuum state $G_{ab}=0$ ($a,b=0,1,2,3,4,5$).
Since we are considering the 3D spatial space in spherical coordinates:
$\vec r \equiv \vec r(r,\theta,\phi)$, here $d\Omega^2 = d\theta^2 + {\rm sin}^2(\theta) d\phi^2$.
Furthermore, the coordinate $N$ is dimensionless and the extra
(space-like) coordinates $\psi$ and $\sigma$ are considered as noncompact.
We define a physical vacuum state on the metric (\ref{ini}) through
the action for a scalar field $\varphi$, which is
nonminimally coupled to gravity
\begin{equation}\label{ac}
I = {\Large\int} d^6x \left[ \frac{^{(6)} {\cal R}}{16\pi G}
+ \frac{1}{2} g^{ab} \varphi_{,a} \varphi_{,b} -
\frac{\xi}{2} {^{(6)} {\cal R}} \varphi^2\right],
\end{equation}
where $^{(6)} {\cal R}=0$ is the Ricci scalar
and $\xi $ gives the coupling of $\varphi$ with gravity.
Implementing the coordinate transformation $N=H t$ and $R=r\psi = r/H$ on the frame
$U^{\psi} = (d\psi / dS)=0$ (considering $H$ as a
constant), followed by the foliation $\psi = H^{-1}$ on the metric (\ref{ini}), we obtain the effective 5D metric
\begin{equation}\label{5d}
^{(5)} dS^2 = dt^2 - e^{2 H t} \left[dR^2 + R^2 d\Omega^2\right] -
d\sigma^2.
\end{equation}
Unfortunately this metric is not Ricci-flat because $^{(5)} {\cal R}= 12 H^2$. However, it becomes Riemann flat in the limit $H %%@
\rightarrow 0$ i.e.
$\left.R^A_{BCD}\right|_{H\rightarrow 0} =0$, describing in this limit
a 5D vacuum given by
\begin{equation}\label{h0}
\left. G_{AB} \right|_{H\rightarrow 0} =0, \qquad (A, B =0,1,2,3,4).
\end{equation}
Thus, keeping this fact in mind, now we consider on the metric (\ref{5d})
the following foliation on the sixth coordinate:
\begin{displaymath}
d\sigma^2 = 2 \Phi(R) \  dR^2.
\end{displaymath}
The effective 4D metric that results is
\begin{equation}\label{4d}
^{(4)} dS^2 = dt^2 - e^{2 H t} \left[ (1+ 2\Phi(R)) dR^2 +
R^2 d\Omega^2\right],
\end{equation}
where $t$ is the cosmic time, $H=\dot a/a$ is the Hubble parameter
for the scale factor $a(t) = a_0 e^{H t}$, with $a_0 = a(t=0)$.
The Einstein equations for the effective 4D metric (\ref{4d}) are
$G_{\mu\nu} = - 8 \pi G \  T_{\mu\nu}$ ($\mu,\nu=0,1,2,3$), where
$T_{\mu\nu}$ is represented by a perfect fluid: $T_{\mu\nu} =
({\rm p} + \rho) u_{\mu} u_{\nu} - g_{\mu\nu} {\rm p}$, being
${\rm p}$ and $\rho$ the pressure and the energy density on the
effective 4D metric (\ref{4d}).
The relevant components of the Einstein tensor are
(the non-diagonal components are zero)
\begin{eqnarray}
G_{tt} & = & = -\frac{\left[ 3H^2 R^2 \left[ 1+4 \Phi (1+\Phi )\right]
+ 2 e^{-2 H t} \left[ R \frac{d\Phi}{d R} + \Phi(1 + 2 \Phi)\right]\right]}{
R^2 \left[ 1+ 2\Phi (R)\right]^2}, \\
G_{RR} & = & \frac{ 2\Phi + 3 H^2 R^2 e^{2 H t} \left[1+2\Phi\right]}{
R^2}, \\
G_{\theta\theta} & = & \frac{R\left[ 3 R H^2 e^{2 H t} \left[
1+ 4\Phi (1+\Phi)\right]\right]}{\left[1+2 \Phi\right]^2}, \\
G_{\phi\phi} & = & \frac{{\rm sin}^2(\theta) R \left[
3 R H^2 e^{2 H t} \left[1+4
\Phi (1+\Phi)\right] + \frac{d\Phi}{dR}\right]}{
\left[1+2 \Phi\right]^2}.
\end{eqnarray}
Now we are aimed to obtain the function $\Phi(R)$.
In order to make that, we are supposing that $\Phi(R)$ is induced
by a mass $M$ located at $R=0$ in absence of expansion ($H=0$),
guaranteeing this way the flatness of (\ref{5d}). Once %%@
we have determined $\Phi(R)$, the Hubble parameter $H$ in (\ref{4d}) is not necessarily null. Thus we
must take the equation
\begin{equation}\label{hh0}
\left.G^t_t\right|_{H\rightarrow 0} =
-\left. 8 \pi G \rho(R,t)\right|_{H\rightarrow 0},
\end{equation}
where $\left.\rho(R,t)\right|_{H\rightarrow 0} = -3M/( 4\pi R^3)$
is the gravitational energy density in absence of expansion.
In this limit, we obtain the following differential equation
for $\Phi(R)$
\begin{equation}\label{df}
- \frac{R^2 \frac{d\Phi}{dR}}{\left[1+2\Phi\right]^2} =
3 G M + \frac{R\Phi}{\left[1+2\Phi\right]}.
\end{equation}
The exact solution for this equation is
\begin{equation}\label{sphi}
\Phi(R) = \frac{-3 G M {\rm ln}(R/R_*)}{R+6 G M {\rm ln}(R/R_*)},
\end{equation}
being $R_*$ the value of $R$ such that $\Phi(R_*)=0$ and $G=M^{-2}_p$
the gravitational constant. Hence, the function $\Phi(R)$
describes the geometrical deformation of the metric induced from a
5D flat metric [the metric (\ref{5d}) with $H=0$], by a mass $M$
located at $R=0$. This function is $\Phi > 0$ (or $\Phi < 0$) for
$R < R_*$ ($R > R_*$), respectively. Furthermore,
$\left.\Phi(R)\right|_{R\rightarrow \infty} \rightarrow 0$ and
thereby the effective 4D metric (\ref{4d}) is (in their 3D
ordinary spatial components) asymptotically flat. In this analysis
we are considering the usual 4-velocities $u^{\alpha} =
(1,0,0,0)$. Thus, for the metric (\ref{4d}), the energy density
and the components of the pressure are
\begin{eqnarray}
&& \rho(R,t) = \frac{3}{8\pi G} \left[ H^2 - \frac{2 GM}{R^3} e^{-2Ht}\right],\\
&& {\rm p}_R(R,t) = -\frac{3}{8\pi G} \left[
H^2 - \frac{2 G M}{R^3} e^{-2H t} {\rm ln}\left(R/R_*\right)\right],\\
&& {\rm p}_{\theta}(R,t) =
-\frac{3}{8\pi G} \left[ H^2 - \frac{GM}{R^3} e^{-2 H t} \left[1-{\rm ln}
\left(R/R_*\right)\right] \right], \\
&& {\rm p}_{\phi}(R,t) = \frac{3}{8\pi G } \left[
H^2 - \frac{GM}{R^3} e^{-2H t}
\left[1-{\rm ln }\left(R/R_*\right)\right]\right].
\end{eqnarray}
Note that ${\rm p}_{\theta} + {\rm p}_{\phi} =0$, so that
the equation of state for a given $R$ is
\begin{equation}\label{state}
\frac{{\rm p}}{\rho} = -1 -\frac{3 G M}{R^3} \frac{\left[
1-{\rm ln}(R/R_*)\right] e^{-2 H t}}{
\left[H^2 - \frac{2 G M}{R^3} e^{-2 H t}\right] },
\end{equation}
being ${\rm p} = {\rm p}_R + {\rm p}_{\theta} + {\rm p}_{\phi}$.
From the equation (\ref{state}) we can see that at the end of inflation,
when the number of e-folds is sufficiently large, the second term
in (\ref{state}) becomes negligible on cosmological scales [on the
infrared (IR) sector], and thereby
\begin{equation}\label{state1}
\left.{\rm p}\right|^{(end)}_{IR} \simeq - \left.\rho\right|^{(end)}_{IR},
\end{equation}
which is the equation of state that describes inflationary cosmology.

The effective 4D action for the universe is
\begin{equation}\label{4}
^{(4)} I = {\Large\int} d^4x \left[\frac{^{(4)} {\cal R}}{16 \pi G}
+ \frac{1}{2} g^{\mu\nu} \varphi_{,\mu} \varphi_{,\nu} -
\frac{\xi_1}{2} \left.^{(4)} {\cal R}\right|_{\Phi=0} \varphi^2 -
\frac{\xi_2(R)}{2} \left.^{(4)} {\cal R}\right|_{H=0} e^{-2 H t} \varphi^2
\right],
\end{equation}
where $^{(4)} {\cal R} = 12 H^2 - (12 G M/ R^3) e^{-2 H t}$
is the effective 4D Ricci scalar for the effective 4D metric (\ref{4d}),
$\xi_1$ gives the coupling of the scalar field $\varphi(\vec R,t)$ with
gravity on the background induced by the foliation of the first extra dimension $\psi$ at
$\Phi(R)=0$ and
$\xi_2(R)$ gives the coupling of $\varphi$ with gravity, on the background induced by
the foliation of the second extra dimension $\sigma$ at $H=0$.
The equation of motion for the field $\varphi$ on the metric (\ref{4d})
is
\begin{eqnarray}
&&\ddot\varphi + 3 H \dot\varphi + e^{-2 H t} \left[
\frac{1}{R^2(1+2\Phi(R))} \frac{\partial}{\partial R}
\left[ R^2 \frac{\partial\varphi}{\partial R} \right] -
\frac{1}{(1+ 2\Phi(R))^2} \frac{\partial\Phi}{\partial R}
\frac{\partial \varphi}{\partial R} \right. \nonumber \\
&& \left.+ \frac{1}{R^2 {\rm sin}(\theta)} \frac{\partial}{\partial\theta}
\left[{\rm sin}(\theta) \frac{\partial\varphi}{\partial\theta}\right]
\right. \nonumber \\
&&+\left. \frac{1}{R^2 {\rm sin}^2(\theta)} \frac{\partial^2\varphi}{\partial\phi^2}
\right] + \left[ \xi_1 \left.^{(4)}{\cal R}\right|_{\Phi=0} +
\xi_2(R) \left.^{(4)} {\cal R}\right|_{H=0} e^{-2 H t}\right]
\varphi =0.\label{5}
\end{eqnarray}
Given the form of $\Phi$ according to (\ref{sphi}), solving this equation is a hard nut to scratch. However, finding solutions in %%@
some limit approximations is easier.

\section{Weak field approximation}

In this section we study the weak field approximation for
the equation (\ref{5}). In that limit approximation the
function $\Phi(R)$ can be written as
\begin{equation}\label{6}
\Phi(R) \simeq - \frac{3 GM}{R} + \left(\frac{6 G M}{R}\right)^2,
\end{equation}
being $M$ the mass of the compact object located at $R=0$. Note that
the function (\ref{6}), as well as the exact one(\ref{sphi}), goes to zero
at $R\rightarrow \infty$. For $M>0$  there is a stable equilibrium for test
particles at $R_{c}=12 G M$ and exhibits a gravitational %%@
repulsion (antigravity) for $R< R_{c}$. Hence, this object have the properties of a white hole \cite{Repulsons}.
In order to obtain solutions of the equation (\ref{5}) we propose
$\varphi(\vec R,t) \sim \varphi_{t}(t) \varphi_R(R)
\varphi_{\theta,\phi}(\theta, \phi)$.
With this choice and using the equation (\ref{df}), we obtain
\begin{eqnarray}
&& \ddot\varphi_t + 3 H \varphi_t - 12 H^2 \xi_1 \varphi_t = -\alpha_t
\varphi_t
e^{-2 H t}, \label{1a} \\
&& \frac{\partial}{\partial R} \left[ R^2 \frac{\partial\varphi_R}{
\partial R} \right] + \left[ 3 G M (1+2\Phi) + R\Phi\right]
\frac{\partial \varphi_R}{\partial R} =
\varphi_R \left[ \frac{12 G M}{R} \xi_2(R) -\alpha_t R^2 + \alpha_R\right]
(1+2\Phi), \label{1b} \\
&& {\rm sin}(\theta) \frac{\partial}{\partial\theta}
\left[ {\rm sin}(\theta) \frac{\partial \varphi_{\theta,\phi}}{
\partial\theta}\right] + \frac{1}{{\rm sin}^2(\theta)} \frac{
\partial^2\varphi_{\theta,\phi}}{\partial\phi^2} = \alpha_R
\varphi_{\theta,\phi}, \label{1c}
\end{eqnarray}
where $\alpha_t$ and $\alpha_R$ are separation constants.

\subsection{confined field}

As a first step we shall study the
initial state of the universe with purely gravitational
energy density at $t=0$. In this case the field $\varphi$ is
confined to negative eigenvalues of energy, i.e. $\alpha_t = -|\alpha_t|$.

The solution of the equation (\ref{1c}) for $\alpha_R=l(l+1)$ is given
by the spherical harmonics $Y_{l,m}(\theta,\phi)$
\begin{equation}
\varphi_{\theta,\phi}(\theta,\phi)\sim \sum_{l,m} A_{lm} \
 Y_{l,m}(\theta,\phi) = \sum_{l,m} A_{lm} \
\sqrt{\frac{(2l+1)(l-m)!}{4\pi (l+m)!}} \  {\cal P}^m_l({\rm
cos}(\theta),
\end{equation}
where $m=-l,-(l-1),...,0,...,(l-1),l$ is a separation constant and
${\cal P}^m_l({\rm cos}(\theta)$ are the Legendre polynomials:
$P^m_l(x) = [(-1)^m / (2^l l!)] (1-x^2)^{m/2} (d^{l+m}/
dx^{l+m}) (x^2-1)^l$.

To complete the study of the spatial dependence of $\varphi(\vec R,t)$
we must solve the equation (\ref{1b}). For solving this equation we propose
\begin{equation}\label{varr}
\varphi_R(R) \sim e^{-\int f(R) dR} \  \varphi_H(R),
\end{equation}
with $f(R) = -[1/ (2 R^2)] \left[3 G M (1+2\Phi(R))+R \Phi(R)\right]$ and thus equation (\ref{1b}) can be replaced by the system
\begin{eqnarray}
&& \frac{d^2\varphi_H}{dR^2} + \frac{2}{R} \frac{d\varphi_H}{dR}+
\left\{\frac{|\alpha_t| R_*}{2 R} - |\alpha_t| - \frac{l(l+1)}{R^2}\right\}
\varphi_H(R) =0,  \label{2a} \\
&& \frac{d\varphi_H}{dR} \left[6 G M(1+2\Phi)+2 R \Phi\right]
+ \varphi_H \left\{ \frac{27 G^2 M^2}{4R^2} (1+2\Phi)^2 +
\frac{3}{4} \Phi^2 \right. \nonumber \\
&& + \left. \frac{9 G M}{2R} \Phi(1+2\Phi) + 3 G M \frac{d\Phi}{dR} +
\frac{\Phi}{2} + \frac{R}{2} \frac{d\Phi}{dR} - 2\Phi |\alpha_t| R^2 -
2\Phi l(l+1) \right. \nonumber \\
&& \left. - \frac{12 G M \xi_2(R)}{R} (1+2\Phi) - \frac{6 G M}{R}
|\alpha_t|\right\} =0.   \label{2b}
\end{eqnarray}
The solution for the equation (\ref{2a}) is
\begin{equation}
\varphi_H(R) = A_{nl} \left(\frac{2R}{\lambda_p}\right)^l \  e^{-\left[
\frac{R}{\lambda_p} + \left(\frac{n \lambda_p}{2R}\right)^2 \left(
1+ \frac{16 n \lambda_p}{3 R}
\right)\right]} \  {\cal L}^{2l+1}_{n+l} \left[\frac{2R}{\lambda_p}\right],
\end{equation}
where ${\cal L}^{2l+1}_{n+l} \left[{2R\over \lambda_p}\right]$ are the
associated Laguerre polynomials with $l=0,1,...,n-1$, for a given
$n$, $A_{nl}$ is a normalization constant, and
\begin{equation}
|\alpha_t| = G^{-1} = \lambda^{-2}_p, \qquad R_* = 4 n \lambda_p,
\qquad M\equiv M_n = \frac{n}{3\lambda_p},
\end{equation}
being $\lambda_p = M^{-1}_p$ the Planckian wavelength.
Hence, the mass of the compact object $M_n=n M_p/3$
and the function $\Phi(R) \equiv \Phi_n(R) = -
(3 G M_n/ R) + \left(6 G M_n/ R\right)^2$ are now quantized.
The interesting
of this result is that for any value of $n$ ($n$ can take any integer
positive value), the eigenvalue $|\alpha_t| = G^{-1} =M^2_p$ is the same.
In other words $|\alpha_t|$ is degenerated.
Note that $\left.\varphi_H(R)\right|_{R\rightarrow 0,R\rightarrow \infty}
\rightarrow 0$, so that it is well defined along $R \geq 0$.
On the other hand, the function $f(R)$ now depends on $n$:
$f_n(R) = -[1/(2 R^2)]
\left[3 G M_n(1+2\Phi_n(R))+ R\Phi_n(R)\right]$.
From equation (\ref{2b}) we obtain that the coupling $\xi_2(R) \equiv
\xi^{n,l}_2(R)$ is given by
\begin{eqnarray}
\xi^{n,l}_2(R) & = & \frac{1}{\varphi_H} \frac{d\varphi_H}{dR}
\left[\frac{R}{2} + \frac{R^2 \Phi_n}{6 G M_n(1+2\Phi_n)}\right] -
\left[\frac{R^3\left(f^2_n - \frac{df_n}{dR}\right)}{12 G M_n (1+
2\Phi_n)} \right. \nonumber \\
 &+ & \left.\frac{2\Phi_n R\left(|\alpha_t|R^2 - l(l+1)\right)-2 R^2 f_n
- R \Phi_n f_n}{12 G M_n (1+2\Phi_n)} - \frac{f_n R}{4}\right],
\end{eqnarray}
which depends on $n$ and $l$.
Furthermore, the solution for the equation (\ref{1a})
for the confined case is [$A_1$ and $A_2$ are integration constants]
\begin{equation}\label{vart}
\varphi_t(t) = e^{-\frac{3 H t}{2}} \left\{ A_1 {\cal K}_{\nu}
\left[\frac{M_p}{H}
e^{-H t}\right]
+ A_2 {\cal I}_{-\nu} \left[\frac{M_p}{H} e^{-H t}\right]\right\},
\end{equation}
where ${\cal K}_{\nu}$ and ${\cal I}_{-\nu}$ are the modified Bessel
functions and $\nu =(1/2)\sqrt{9+ 48 \xi_1}$.
Hence, the solution for $\varphi(\vec R, t)$ in the case where
the field is confined, will be
\begin{equation}
\varphi(\vec R,t) =   \varphi_t(t)
\sum^{n-1}_{l=0} \sum^{l}_{m=-l} \  \varphi_R(R)
\left[ A_{lm} Y_{l,m}(\theta,\phi)
+ A^{\dagger}_{lm} Y^*_{l,m}(\theta,\phi)\right],
\end{equation}
with $\varphi_R(R)$ and $\varphi_t(t)$ given respectively by
 (\ref{varr}) and (\ref{vart}). Note that $\varphi(\vec R,t)$
is well defined for all $R \geq 0$. \\

\subsection{Dispersive case}

Now we study the dispersive case ($\alpha_t = |\alpha_t|=k^2_R$)
for $\varphi(\vec R,t)$. In this case $k_R$ is the wavenumber related to the coordinate $R$. Proceeding in a similar manner that %%@
in the previous section for solving the equation (\ref{1b}), we propose
\begin{equation}
\varphi_R(R) = \varphi_{k_R}(R) \  e^{-\int f_n(R) dR},
\end{equation}
such that $\varphi_{k_R}(R) = \varphi_R[\Phi=0]$. With this choice the expression (\ref{1b}) can be replaced by the equations
\begin{eqnarray}
&& \frac{d^2\varphi_{k_R}}{dR^2} + \frac{2}{R} \frac{d\varphi_{k_R}}{dR} +
\varphi_{k_R} \left[k^2_R - \frac{l(l+1)}{R^2}\right]=0, \label{3a} \\
&& \frac{d\varphi_{k_R}}{dR} \left[6 G M_n(1+2\Phi_n) + 2R\Phi_n\right]
+\varphi_{k_R} \left[ R^2 \left(f^2_n - \frac{df_n}{dR}\right) - 2 R f_n
+ 2\Phi_n \left(k^2_R R^2 - l(l+1)\right) \right.\nonumber \\
&& - \left. 3 G M_n (1+2\Phi_n) f_n -
R \Phi_n f_n - \frac{12 G M_n \xi_2(R)}{R} \left(1+2\Phi_n \right)\right]=0.
\label{3b}
\end{eqnarray}
The solution for the equation (\ref{3a}) is
\begin{equation}\label{3c}
\varphi_{k_R}(R) =
\frac{A}{\sqrt{R}} \  {\cal J}_{l+1/2}[k_R R],
\end{equation}
where ${\cal J}_{l+1/2}$ are the Bessel functions.
On the other hand, from the equation (\ref{3b}) we obtain the coupling
$\xi^{n,l}_2(R)$
\begin{eqnarray}
\xi^{n,l}_2(R) & = & \frac{R}{12 G M_n (1+2\Phi_n)} \left\{
\frac{1}{\varphi_{k_R}}\frac{d\varphi_{k_R}}{dR} \left[ 6 G M_n
(1+2\Phi_n) + 2 R \Phi_n\right] - \left[ R^2\left(f^2_n -
\frac{df_n}{dR}\right)
\right. \right. \nonumber \\
&-& \left.\left. 2 R f_n + 2\Phi_n \left[k^2_R R^2 - l(l+1)\right]
-3 G M_n \left( 1+2\Phi_n \right) f_n - R \Phi_n
f_n\right]\right\},
\end{eqnarray}
with $f_n(R) = -[1/ (2R^2)] \left[3 G M_n \left(1+ 2\Phi_n(R)\right)
+ R \Phi_n(R)\right]$.
The solution of the equation (\ref{1a}) for the dispersive case is
\begin{equation}                  \label{26}
\varphi_t(t) \sim \xi_{k_R}(t)
= A \  e^{-3 H t/2} {\cal H}^{(2)}_{\nu}\left[\frac{k_R}{H} e^{-H t}\right],
\end{equation}
being $A$ a constant, $\nu = (1/2)\sqrt{9 + 48 \xi_1}$ and
${\cal H}^{(2)}_{\nu}$ is the second kind Hankel function.
Taking into account all the possible values of $k_R$, the
complete solution for the field $\varphi(\vec R,t)$ can be
written as
\begin{eqnarray}
\varphi(\vec R, t) &=& \frac{1}{(2\pi)^{3/2}}
e^{-\left[\left(\frac{n\lambda_p}{2R}\right)^2
\left(1+\frac{16 n \lambda_p}{3R}\right)\right]}
{\Large\int} d^3 k_R \sum^{l}_{m=-l} \sum^{n-1}_{l=0} \left[
a_{k_R l m} Y_{l,m}(\theta,\phi) \varphi_{k_R}(R)
\xi_{k_R}(t) \right. \nonumber \\
&+& \left. a^{\dagger}_{k_R l m} Y^*_{l,m}(\theta,\phi)
\varphi^*_{k_R}(R) \xi^*_{k_r}(t)\right], \label{var}
\end{eqnarray}
where $\varphi_{k_R}(R)$ and $\xi_{k_R}(t)$ are given respectively
by the expressions (\ref{3c}) and (\ref{26}).
Notice that the exponential factor in (\ref{var}) tends to $0$ for $R/\lambda_p
\rightarrow 0$. It solves all possible UV divergences of the
$\varphi$-fluctuations in a natural manner.\\

\subsection{Large scale asymptotic behavior}

In general, the spatial homogeneity and isotropy of the universe on cosmological scales is an accepted fact. This property %%@
introduce new symmetries to the solution (\ref{var}) allowing to explore the possibility to recover the well-known nearly  %%@
invariant spectrum on super-Hubble scales, usually obtained
from inflationary models. Thereby it results convenient to study the %%@
large scale asymptotic behavior of the solution (\ref{var}). We can obtain this limit approximation by considering the conditions  %%@
$R/\lambda_p > R H \gg 1$ and $k_R R \gg 1$ (the IR sector). Thus, taking
into account that $\left.{\cal J}_{\mu}[x]\right|_{x\gg 1} \simeq
\sqrt{2 / (\pi x)}\,
{\rm cos}\left[x-\mu\pi/ 2 - \pi/ 4\right]$, the expression (\ref{var}) on
the infrared (IR) sector becomes
\begin{eqnarray}
\left.\varphi(\vec R,t)\right|_{IR} & \simeq & \frac{1}{(2\pi)^{3/2}}
\sqrt{\frac{
2}{\pi}} \frac{1}{R} {\Large\int} \frac{d^3 k_R}{\sqrt{k_R}}
\sum^{l}_{m=-l} \sum^{n-1}_{l=0} \left[ a_{k_R l m}
Y_{l,m}(\theta,\phi) e^{i[k_R R - \frac{l\pi}{2}]}\left.\xi_{k_R}(t)
\right|_{IR}\right. \nonumber \\
& + & \left. a^{\dagger}_{k_R l m} Y^*_{l,m}(\theta,\phi)
e^{-i[k_R R - \frac{l\pi}{2}]}\left.\xi^*_{k_R}(t)\right|_{IR}\right].
\label{30}
\end{eqnarray}
On the other hand, it is important to emphasize that on cosmological
scales we can consider $\Phi(R)\simeq 0$.
Considering these limit conditions and transforming (\ref{30}) to
{\em cartesian coordinates}, we obtain
\begin{equation} \label{cam}
\left.\varphi(\vec{\bar{R}},t)\right|_{IR}^{(\Phi (R)\simeq 0)}\simeq\frac{1}{
(2\pi)^{3/2}}\int d^{3}k_{\bar{R}}\left[a_{k_{\bar{R}}}e^{i\vec{k}_{\bar{R}}
\cdot\vec{\bar{R}}}\left.\xi _{k_{\bar{R}}}(t)\right|_{IR}^{(\Phi (R)\simeq 0)}+a_{k_{\bar{R}}}^{\dagger}
e^{-i\vec{k}_{\bar{R}}\cdot\vec{\bar{R}}}\left.\xi _{k_{\bar{R}}}^{*}(t)\right|_{IR}^{(\Phi (R)\simeq 0)}\right],
\end{equation}
being $d\bar{R}^{2}=dx^2+dy^2+dz^2$ with
$x=\bar R \  {\rm sin}(\theta) \  {\rm cos}(\phi)$,
$y=\bar R \ {\rm sin}(\theta) \  {\rm sin}(\phi)$ and
$z=\bar R \  {\rm cos}(\phi)$. In the obtaining of (\ref{cam}) we have used the Rayleigh expansion %%@
$e^{i\vec{k}_{\bar{R}}\cdot\vec{\bar{R}}}=
4\pi \sum _{l,m} i^{l}{\cal J}_{l}(k_{R}R)Y_{l,m}^{*}(\Omega _{p})
Y_{l,m}(\Omega _{R})$ and the relations between annihilation-annihilation and creation-creation
operators from spherical to cartesian representations $a_{k_{R}lm}=i^{l} k_{\bar{R}}\int d\Omega _{k_{\bar{R}}}
Y_{l,m}^{*} (\Omega _{k_{\bar{R}}})a_{k_{\bar{R}}}$ and $a_{k_{R}lm}^{\dagger}=(-i)^{l} k_{\bar{R}}\int
d\Omega _{k_{\bar{R}}}Y_{l,m} (\Omega _{k_{\bar{R}}})
a_{k_{\bar{R}}}^{\dagger}$. Thus, the corresponding asymptotic expression for the modes on IR sector is
\begin{equation}\label{modosIR}
\left.\xi_{k_R}(t) \right|_{IR}^{(\Phi (R)\simeq 0)} \simeq -\frac{i}{2}
\sqrt{\frac{1}{\pi H}} \Gamma(\nu) \left(\frac{k_{\bar R}}{2H}\right)^{-\nu}
e^{H t (\nu-3/2)}.
\end{equation}
Note that the expressions (\ref{cam}) and (\ref{modosIR})
are the same that we have obtained in previous papers \cite{MB1}-\cite{grav1}. In this manner,
we recover cosmological solutions which agree
with an isotropic and homogeneous universe
in absence of the potential $\Phi(R)$ (and therefore of the
compact object).
As in \cite{MB1}-\cite{grav1}, in what follows we shall use the %%@
cartesian coordinate representation in order to make explicit the asymptotic
spatial homogeneity and isotropy on very large scales.

Now, we are able to calculate the effective 4D super-Hubble squared fluctuations,
for the cosmological limit in cartesian coordinates
$\left<\varphi^{2} \right>_{IR}^{(\Phi (R)\simeq 0)}$, which are given by
\begin{equation} \label{asim4}
\left<\varphi^2\right>_{IR}^{(\Phi (R)\simeq 0)}
=\frac{1}{2\pi^2}\int _{0}^{ k_{H}(t)}
\frac{ dk_{\bar{R}} }{k_{\bar{R}} } k_{\bar{R}}^{3}\left.
\left[\xi_{k_{\bar{R}}}(t) \xi_{k_{\bar{R}}}^{*}(t)
\right]\right|_{IR}^{(\Phi (R)\simeq 0)},
\end{equation}
where $k_H=\sqrt{12\xi _{1}+(9/4)}\,H  e^{H t}$ is the Hubble wavenumber on physical coordinates
in a de Sitter expansion. Inserting (\ref{modosIR}) into (\ref{asim4}) we obtain
\begin{equation}\label{asim5}
\left<\varphi^2\right>_{IR}^{(\Phi (R)\simeq 0)}=\frac{2^{2\nu -3}}{\pi^3}\Gamma^{2}(\nu)
H^{2\nu -1}e^{-(3-2\nu)Ht}\int _{0}^{ k_{H}}
\frac{dk_{\bar{R}}}{k_{\bar{R}}} k_{\bar{R}}^{3-2\nu}.
\end{equation}
Note that when the coupling parameter $\xi_{1}\ll 1$,
the index $\nu\simeq 3/2$
and thus the corresponding 3D power spectrum ${\cal P}
\left( k_{\bar{R}}\right) \sim k_{\bar{R}}^{3-2\nu}$
becomes nearly scale invariant. Performing the
remaining integration, the expression (\ref{asim5}) yields
\begin{equation}\label{asim6}
\left<\varphi^2\right>_{IR}^{(\Phi (R)\simeq 0)}=\frac{2^{2\nu -3}}{3-2\nu}
\frac{\Gamma^{2}(\nu)}{\pi^3}H^{2\nu -1}e^{-(3-2\nu)Ht}
k_{H}^{3-2\nu}
=\frac{2^{2\nu-3} H^2
\Gamma^2(\nu)}{\pi^3 (3-2\nu)\left(12\xi _{1}+\frac{9}{4}\right)^{\nu-3/2}}.
\end{equation}
The 3D power spectrum corresponds to the spectral index
$n_{s}=4-\sqrt{9+48 \xi_{1}}$. On the other hand, it is well known from observational evidences \cite{OBS1} that $n_{s}=0.97 \pm %%@
0.03$.
Hence, we can establish the next range of values for the coupling parameter $0\le \xi_{1}\leq 75.75\times 10^{-4}$.

\subsection{Small Scale Approximation}

\begin{center}
\begin{figure}
\epsfysize=9cm
\epsfbox{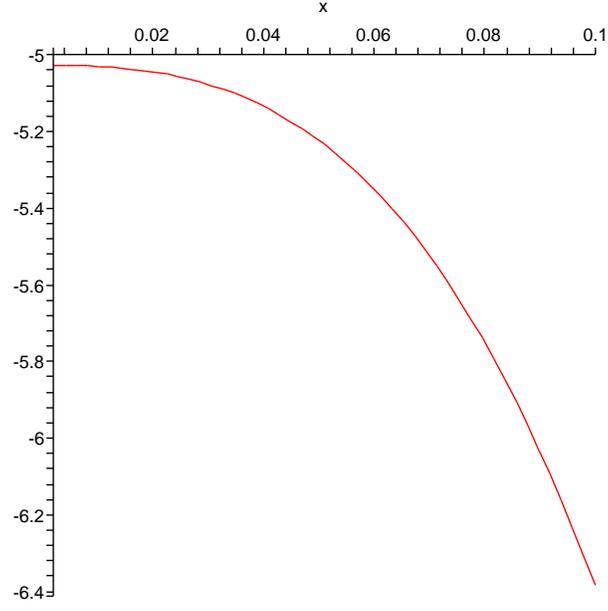}
\caption{\label{f1} ${\rm
log}_{10}\left[{\left.\delta\rho(k_R)\right|_{SS} \over
\rho}\right]$ as a function of $x=k_R/k_p$. The energy density
fluctuations are nearly scale invariant on large scales, but
become scale dependent on the blue sector.}
\end{figure}
\end{center}

Once we have studied the asymptotic behavior on large scale of the scalar
field (\ref{var}), we are aimed to obtain its behavior on
small scales $k_{H}R\ll 1$. Since the modes responsible for the posterior structure formation must be those that become classical
on super-Hubble scales after inflation ends, in order to
obtain the corresponding spectrum on small scales, we should consider
those modes whose physical length scale is a little bigger than the
horizon scale (the Hubble radius) at that time. Hence, for
small scales, in presence of the compact object [and thus for $\Phi (R)
\neq 0$], we must understand the today astrophysical
scales ($\sim$ 100 Mpc). In our conventions we will refer to this part of the spectrum as the small scale (SS) spectrum.
Therefore, the  effective small scale squared fluctuations in
presence of $\Phi(R)\neq 0$ are given by
\begin{equation}\label{sma3}
\left<\varphi^{2}\right>_{SS}^{(\Phi(R)\neq 0)}=\frac{1}{2\pi^2}
e^{-\left[\frac{1}{2}\left(\frac{n\lambda _{p}}{R}\right)^2\left(1+\frac{16n\lambda _{p}
}{3 R}\right)\right]} \int _{\epsilon k_{H}}^{k_{H}}\frac{dk_{R}}{k_{R}}k_{R}^{3}\left.\left[\xi _{k_R}\xi %%@
_{k_R}^{*}\right]\right|_{IR}^{(\Phi (R)\simeq 0)}
\end{equation}
where $\epsilon =(k_{max}^{(IR)}/k_p)\ll 1$ is a dimensionless constant
parameter, and
$k_{max}^{IR}=\left.k_{H}(t_{i})=\sqrt{12\xi _{1}+(9/4)}H e^{Ht}
\right|_{t=t_i}$ is the
wavenumber related with the Hubble radius at the
time $t_i$ when the horizon re-enters and $k_{p}$ is the
Planckian wavenumber.
Inserting the expression (\ref{modosIR}) into (\ref{sma3}) we obtain
\begin{equation}\label{sma4}
\left<\varphi^{2}\right>_{SS}^{(\Phi(R)\neq 0)}=\frac{2^{2\nu-3}}{3-2\nu}\frac{\Gamma^{2}(\nu)}{\pi^3}
H^{2\nu-1}e^{-(3-2\nu)Ht}
e^{-\left[\frac{1}{2}\left(\frac{n\lambda _{p}}{R}\right)^2\left(1+\frac{16n\lambda _{p}
}{3 R}\right)\right]} k_{H}^{3-2\nu}(1-\epsilon)^{3-2\nu}.
\end{equation}
Comparing (\ref{asim6}) with (\ref{sma4}), we obtain
\begin{equation}\label{sma5}
\left<\varphi^{2}\right>_{SS}^{(\Phi(R)\neq 0)}=
e^{-\left[\frac{1}{2}\left(\frac{n\lambda _{p}}{R}\right)^2\left(1+\frac{16n\lambda _{p}
}{3 R}\right)\right]}
(1-\epsilon)^{3-2\nu}
\left<\varphi^{2}\right>_{IR}^{(\Phi (R)\simeq 0)}.
\end{equation}
However, since $\epsilon \ll 1$ then we can make
the approximation $(1-\epsilon)^{3-2\nu}\simeq 1$. Hence, the expression  (\ref{sma5}) becomes
\begin{equation}
\left<\varphi^{2}\right>_{SS}^{(\Phi(R)\neq 0)}
\simeq
e^{-\left[\frac{1}{2}\left(\frac{n\lambda _{p}}{R}\right)^2\left(1+\frac{16n\lambda _{p}
}{3 R}\right)\right]}
\left<\varphi^2\right>_{IR}^{(\Phi (R)\simeq 0)}.
\end{equation}
Finally, we can write the power spectrum of $\left<\varphi^2\right>_{SS}$,
making $k_R = 2\pi/ R$ and $k_p = 2\pi / \lambda_p$
\begin{equation}
\left.{\cal P}(k_R)\right|_{\left<\varphi^2\right>_{SS}} \sim
e^{-\left[\frac{1}{2}\left(n \frac{k_R}{k_p}\right)^2
\left(1+\frac{16n}{3}\frac{k_R}{k_p}\right)\right]}
k^{3-2\nu}.
\end{equation}
This result resembles the whole obtained by J. Einasto
{\em et. al}\cite{einasto}, who obtained from observation that the spectrum
of galaxies at present is determined by two fundamental spectral indices:
$n_s \simeq 1$ (which is dominant on cosmological scales) and
$m \simeq -1.9$ (which is dominant on astrophysical scales).
Such a result is in agreement with our calculations.
In the figure (\ref{f1}) we have plotted
${\rm log}_{10}\left[{\left.\delta\rho(k_R)\right|_{SS}
\over \rho}\right]$, where
\begin{displaymath}
\frac{\left.\delta\rho(k_R)\right|_{SS}}{\rho} \simeq
\frac{\xi_1}{2} \  {^{(4)} {\cal R}} \frac{\left<\varphi^2
\right>^{(\Phi(R)\simeq 0)}_{SS}}{V(\left<\varphi\right>)},
\end{displaymath}
as a function of $x=k_R/k_p$. We have used $H=1.7 \times 10^{-7} \
{\rm M_p}$ and $V(\left<\varphi\right>) = {\xi_1\over 2} {^{(4)}
{\cal R}} \left<\varphi\right>^2_{IR}$ $ = 10^{-20} \  {\rm
M^4_p}$, $\xi_1 = 75 \times 10^{-4}$ and $n=10$ with $\nu = 1.518$
(i.e., for a spectral index $n_s = 0.964$). Note that the energy
density fluctuations are scale dependent on scales of the order of
$k_R/k_p > 10^{-2}$, which corresponds to $R < 10^{-3} H^{-1}$
during inflation.

\section{Final Comments}

In this letter using ideas of STM theory of gravity we have studied an effective 4D inflationary expansion
of the universe that emerges from a 6D vacuum state. Under this approach, the
expansion is affected by a geometrical deformation
induced by the gravitational attraction of a white hole of
mass $M_n =n M_p/3$. This deformation is described by the function
$\Phi_n(R)$, which tends to zero on cosmological scales. However,
on sub Hubble scales this function is determinant for the spectrum of
the squared $\varphi$-fluctuations. An important result of our
formalism is that the spectrum of energy density fluctuations
is almost independent of the scale on cosmological
scales, but this is not the case on current small (sub Hubble) scales.
We suggest that it could be the origin of the current $k_R$-non invariant
spectrum  observed for the distribution of galaxies on astrophysical scales.
A different, but in some sense emparented approach, was recently
suggested in the literature\cite{chino}. For simplicity, our calculations
were made in the weak field approximation of the expression (\ref{6}) for
$\Phi(R)$, so that they are valid on the infrared and blue sectors
of the spectrum. However, these  results could be extensive for strong
interactions, using the exact expression (\ref{sphi}) for $\Phi(R)$.
Notice that in this letter we have neglected
the last term in the action (\ref{4})
to obtain the  matter density spectrum.
This term, through
$\xi_2(R)$, could be responsible for the present day galaxy clusters
correlations\cite{einasto}. A more exhaustive study of this topic
will be done in a forthcoming paper.

\vskip .2cm
\centerline{\bf{Acknowledgements}}
\vskip .2cm
JEMA acknowledges CNPq-CLAF and UFPB
and MB acknowledges CONICET
and UNMdP (Argentina) for financial support.\\

%\begin{center}
%\begin{figure}
%\epsfig{file=fgi38.eps,width=9cm, angle=0} \vspace{0cm} \noindent
%\caption{\label{f1} ${\rm
%log}_{10}\left[{\left.\delta\rho(k_R)\right|_{SS} \over
%\rho}\right]$ as a function of $x=k_R/k_p$. The energy density
%fluctuations are nearly scale invariant on large scales, but
%become scale dependent on the blue sector.}
%\end{figure}
%\end{center}
\vskip 10cm

\end{document}